\documentclass[12pt,spanish,english]{article}
\usepackage[T1]{fontenc}
\usepackage[latin9]{inputenc}
\usepackage{geometry}
\usepackage{amsmath}
\usepackage{amssymb}
\usepackage{setspace}
\usepackage{esint}
\makeatletter
\newcommand{\lyxaddress}[1]{
\par {\raggedright #1
\vspace{1.4em}
\noindent\par}
}

\makeatother

\usepackage{babel}
\addto\shorthandsspanish{\spanishdeactivate{~<>}}

\begin{document}

\title{Relativistic heat flux for a single component charged fluid in the
presence of an electromagnetic field}

\author{A. L. Garcia-Perciante$^{1}$, A. Sandoval-Villalbazo$^{2}$, D.
Brun-Battistini$^{2}$}

\maketitle

\lyxaddress{$^{1}$Depto. de Matematicas Aplicadas y Sistemas, Universidad Autonoma
Metropolitana-Cuajimalpa, Prol. Vasco de Quiroga 4871, Mexico D.F
05348, Mexico.}

\lyxaddress{$^{2}$Depto. de Fisica y Matematicas, Universidad Iberoamericana
Ciudad de Mexico, Prolongacion Paseo de la Reforma 880, Mexico D.
F. 01219, Mexico.}
\begin{abstract}
Transport properties in gases are significantly affected by temperature.
In previous works it has been shown that when the thermal agitation
in a gas is high enough, such that relativistic effects become relevant,
heat dissipation is driven not solely by a temperature gradient but
also by other vector forces \cite{israel63}-\cite{benedicks}. In
the case of relativistic charged fluids, a heat flux is driven by
an electrostatic field \emph{even in the single species case} \cite{benedicks}.
The present work generalizes such result by considering also a magnetic
field in an arbitrary inertial reference frame. The corresponding
constitutive equation is explicitly obtained showing that both electric
and magnetic forces contribute to thermal dissipation. This result
may lead to relevant effects in plasma dynamics.
\end{abstract}

\section{Introduction}

Linear irreversible thermodynamics predicts linear relations between
dissipative fluxes and thermodynamic forces. These constitutive, or
closure, equations when generalized for high temperature fluids feature
cross-like effects. In particular, it has been shown that a heat flux
is driven by an external electrostatic field in a charged single component
fluid. This corresponds to an electro-thermal effect in simple gases
which is not present in the non-relativistic, low temperature, case
when the full Boltzmann equation is considered \cite{benedicks}.
In the present paper such an effect is generalized to the electromagnetic
case considering an additional magnetic field and working in an arbitrary
reference frame.

Boltzmann's equation has been applied by other authors to address
both the single component charged gas as well as the mixture in both
relativistic and non-relativistic scenarios. In particular, Spitzer
established a heat flux depending on the electrostatic field for a
monocomponent fluid by neglecting the time derivative term, assuming
a steady state situation \cite{spitzer-1,spitzer1}. On the other
hand, non-relativistic two component plasmas have been addressed widely
in the literature where electro-thermal effects correspond to cross
effects that rely on the system being a mixture \cite{Marshall,balescu,key-2}.
Kremer et. al. adressed the relativistic charged mixture obtaining
Fourier and Ohm's law as well as electro-thermal effects driven by
the electric field \cite{kremerohm}.

The formalism here presented follows the standard steps in the Chapman-Enskog
method whithin Marle's relaxation approximation for a charged single
relativistic fluid. Once the electromagnetic contribution to the non-equilibrium
part of the distribution is established, the heat flux is calculated
based on the expression in terms of the energy flux arising solely
from the chaotic component of the velocities \cite{microscopic}.
The result obtained consists on a general expression for the heat
flux present in a simple fluid driven by an external electromagnetic
field in an arbitrary reference frame.

The rest of the paper is divided as follows. Section II is devoted
to the description of the fundamental elements of relativistic kinetic
theory where Boltzmann's equation is written for a charged single
component fluid and the corresponding transport equations are obtained.
A simplified collision model is incorporated in Sect. III in order
to establish the basic structure of the non-equilibrium part of the
distribution function. The heat flux is obtained in Sect. IV, where
the explicit constitutive equation is shown and the corresponding
transport coefficient calculated. The explicit dependence of the heat
flux on the electromagnetic force is also shown. Conclusions and final
remarks are included in Sect. V.

\section{Transport equations for a relativistic charged simple fluid}

Relativistic molecular dynamics become relevant in gases when thermal
agitation is such that the particle's chaotic velocities become close
to the speed of light. The relativistic nature of the gas is measured
through the relativistic parameter $z=kT/mc^{2}$. When $z\gtrsim1$,
the thermal energy is comparable to the rest mass of the individual
molecules and relativistic corrections to particle dynamics permeate,
through statistical averages, to measurable corrections in state variables
and their evolution equations.

From the kinetic theory point of view, the formal treatment of high
temperature gases is based on the relativistic Boltzmann equation.
For the case of a single charged relativistic fluid, such equation
can be written as \cite{de groot6,cercigniani}
\begin{equation}
v^{\alpha}\frac{\partial f}{\partial x^{\alpha}}+\frac{q}{m}F^{\alpha\nu}v_{\nu}\frac{\partial f}{\partial v^{\alpha}}=J\left(ff'\right)\label{eq:1-1}
\end{equation}
Here $f$ is the relativistic distribution function which, as in the
non relativistic case, yields the occupation number of phase-space
cells with volume $d^{3}xd^{3}v$. As long as the fluid is dilute
and no intense gravitational fields are present, a flat Minkowski
metric $\eta^{\mu\nu}$ (with a +++- signature) is appropriate where
$x^{\mu}=\left[\vec{r},ct\right]$ and $v^{\mu}=\gamma_{w}\left[w^{\mu},c\right]$,
with $\gamma_{w}=\left(1-\frac{w^{\ell}w_{\ell}}{c^{2}}\right)^{-1/2}$
being the Lorentz factor. Here and in the rest of the work the Einstein
summation convention is used, with greek indices running from 1 to
4 and latin ones up to 3. Also, in Eq. (\ref{eq:1-1}) the external
force is expressed in terms of the electromagnetic field tensor $F^{\alpha\nu}$.
Notice that in the electromagnetic case, the inclusion of a velocity
dependent force field does not alter the streaming side of the equation
because of the antisymmetry properties of $F^{\alpha\nu}$, that is
\begin{equation}
\frac{\partial}{\partial v^{\alpha}}\left(F^{\alpha\nu}v_{\nu}\right)=F^{\alpha\nu}\frac{\partial v_{\nu}}{\partial v^{\alpha}}=F^{\alpha\nu}\delta_{\nu\alpha}=0\label{eq:2-1}
\end{equation}

Conserved quantities and the expressions for dissipative fluxes and
state variables are obtained following a similar procedure as the
one carried out in Ref. \cite{microscopic}. Starting from Boltzmann's
equation (Eq. (\ref{eq:1-1})) one multiplies both sides by a collisional
invariant $\psi\left(v^{\mu}\right)$ and integrate over velocity
space, which yields 
\begin{equation}
\frac{\partial}{\partial x^{\alpha}}\int\psi\left(v^{\mu}\right)v^{\alpha}fd^{*}v-\int F^{\alpha\nu}v_{\nu}\psi\left(v^{\mu}\right)\frac{\partial f}{\partial v^{\alpha}}d^{*}v=\int J\left(ff'\right)d^{*}v\label{eq:4}
\end{equation}
where the fact that velocity and position are independent variables
in phase space has been used for the first term on the left hand side.
The invariant volume element is given by $d^{*}v=\gamma_{w}^{-1}d^{3}v$
\cite{Heat}. For the second term on the left hand side of Eq. (\ref{eq:4})
one has
\begin{equation}
\int F^{\alpha\nu}v_{\nu}\psi\left(v^{\mu}\right)\frac{\partial f}{\partial v^{\alpha}}d^{*}v=\int\frac{\partial}{\partial v^{\alpha}}\left[F^{\alpha\nu}v_{\nu}\psi\left(v^{\mu}\right)f\right]d^{*}v-\int f\psi\left(v^{\mu}\right)\frac{\partial}{\partial v^{\alpha}}\left[F^{\alpha\nu}v_{\nu}\right]d^{*}v-\int fF^{\alpha\nu}v_{\nu}\frac{\partial}{\partial v^{\alpha}}\left[\psi\left(v^{\mu}\right)\right]d^{*}v\label{eq:5}
\end{equation}
where the first term vanishes using Gauss' theorem and the fact that
the distribition function vanishes (exponentially) on the boundaries
of the $v$ domain. The second term also vanishes because of the antisymmetry
of the field tensor. For the right hand side, the symmetry properties
of the collision operator, together with the fact that $\psi\left(v^{\mu}\right)$
is a collisional invariant, leads to the vanishing of such term and
thus one obtains
\begin{equation}
\frac{\partial}{\partial x^{\alpha}}\int\psi\left(v^{\alpha}\right)v^{\alpha}fd^{*}v-\frac{q}{m}\int fF^{\alpha\nu}v_{\nu}\frac{\partial\psi\left(v^{\beta}\right)}{\partial v^{\alpha}}d^{*}v=0\label{6}
\end{equation}
Considering $\psi\left(v^{\mu}\right)=1$ in Eq. (\ref{6}) one has
\begin{equation}
\frac{\partial N^{\alpha}}{\partial x^{\alpha}}=0\label{7}
\end{equation}
with $N^{\alpha}=\int v^{\alpha}fd^{*}v$ being the particle four
flux. The relation of this flux with the hydrodynamic four velocity
$\mathcal{U}^{\alpha}$ follows as in the neutral case as 
\begin{equation}
N^{\alpha}=n\mathcal{U}^{\alpha}\label{8}
\end{equation}
and thus, the conservation of such quantity leads to the continuity
equation namely,
\begin{equation}
\mathcal{U}^{\mu}\frac{\partial n}{\partial x^{\mu}}=-n\frac{\partial\mathcal{U}^{\mu}}{\partial x^{\mu}}\label{eq:17}
\end{equation}
For the momentum-energy invariant $\psi\left(v^{\beta}\right)=mv^{\beta}$
one obtains the following balance equation
\begin{equation}
\frac{\partial T^{\beta\alpha}}{\partial x^{\alpha}}=qn\mathcal{U}_{\nu}F^{\beta\nu}\label{9}
\end{equation}
where the fluid's energy-momentum tensor is defined as
\begin{equation}
T^{\alpha\beta}=\int mv^{\alpha}v^{\beta}fd^{*}v\label{10}
\end{equation}
Also, by expressing this quantity in terms of a 3+1 decomposition
of a symmetric second rank tensor leads to the same expression as
in Ref. \cite{microscopic}, that is
\begin{equation}
T^{\alpha\beta}=\frac{n\varepsilon}{c^{2}}\mathcal{U}^{\alpha}\mathcal{U}^{\beta}+ph^{\alpha\beta}+\frac{1}{c^{2}}q^{\alpha}\mathcal{U}^{\beta}+\frac{1}{c^{2}}q^{\beta}\mathcal{U}^{\alpha}+\Pi^{\alpha\beta}\label{e11}
\end{equation}
where $\varepsilon$ is the internal energy density, $p$ the hydrostatic
pressure, $q^{\alpha}$ the heat flux and $\Pi^{\alpha\beta}$ Navier's
viscous tensor. Also, $h^{\alpha\beta}$ is the so-called spatial
projector which is given by
\begin{equation}
h^{\mu\nu}=\eta^{\mu\nu}+\frac{\mathcal{U}^{\mu}\mathcal{U}^{\nu}}{c^{2}}\label{eq:h}
\end{equation}
The force term on the right hand side of Eq. (\ref{9}) can be expressed
as the four-divergence of the electromagnetic stress tensor $M^{\alpha\beta}$
\cite{cercigniani,jackson} such that the energy-momentum conservation
equation for the single charged fluid can be written as
\begin{equation}
\frac{\partial}{\partial x^{\alpha}}\left(T^{\beta\alpha}+M^{\alpha\beta}\right)=0\label{e12}
\end{equation}
Equation (\ref{e12}) contains both the momentum balance equation
and the total energy balance. In order to write an expression for
the internal energy evolution one projects Eq. (\ref{eq:12}) in the
direction of the hydrodynamic velocity namely
\begin{equation}
\mathcal{U}_{\beta}\frac{\partial}{\partial x^{\alpha}}\left(T^{\beta\alpha}+M^{\alpha\beta}\right)=0\label{eq:dd}
\end{equation}
which yields
\begin{equation}
n\dot{\varepsilon}+p\frac{\partial\mathcal{U}^{\mu}}{\partial x^{\nu}}=\mathcal{U}_{\mu}\frac{1}{c^{2}}\dot{q}^{\mu}-\frac{\partial q^{\nu}}{\partial x^{\nu}}\label{eq:18-1}
\end{equation}
where the dot indicates a fluid's proper time derivative, i.e. $\dot{A}^{\mu}=\mathcal{U}^{\nu}\frac{\partial A^{\mu}}{\partial x^{\nu}}$.
Notice that the electromagnetic field term vanishes since
\begin{equation}
\mathcal{U}_{\mu}\frac{\partial M^{\mu\nu}}{\partial x^{\nu}}=qF^{\mu\nu}\mathcal{U}_{\mu}\mathcal{U}_{\nu}=0\label{eq:3-4}
\end{equation}
where we have used that $F^{\mu\nu}$ is antisymmetric. For the momentum
balance one obtains, from Eq. (\ref{e12}),
\begin{equation}
\left(\frac{n\varepsilon}{c^{2}}+\frac{p}{c^{2}}\right)\dot{\mathcal{U}^{\mu}}+\left(p_{,\alpha}+\frac{1}{c^{2}}\dot{q}_{\alpha}+\pi_{\alpha;\nu}^{\nu}\right)h^{\mu\alpha}+\frac{1}{c^{2}}\left(\mathcal{U}_{;\nu}^{\mu}q^{\nu}+\theta q^{\mu}\right)=qF^{\mu\nu}\mathcal{U}_{\nu}\label{eq:19}
\end{equation}
Equations (\ref{eq:17}), (\ref{eq:18-1}) and (\ref{eq:19}) constitute
the set of transport equations for a relativistic single component
inviscid fluid in the presence of an electromagnetic field . A constitutive
equation that expresses the heat flux in terms of the forces is required
in order to close the set such that it can appropriately describe
the corresponding dynamics. The Chapman-Enskog procedure will be used
in the following sections in order to obtain such a relation.

\section{The electromagnetic contribution to $f^{\left(1\right)}$}

In this section, the Chapman-Enskog method is used in order to establish
a first order in the gradients correction to the equilibrium distribution
function using Eq. (\ref{eq:1-1}). For the system here considered,
all gradients of the hydrodynamic velocity are negelcted since the
heat flux, being a vector flux, will not be coupled to second rank
tensor forces. 

For the system in hand, a simple charged gas in the presence of an
electromagnetic field, the acceleration acting on the molecules that
appears in the second term of Eq. (\ref{eq:3-2}) is calculated using
the electromagnetic field tensor which, in cartesian coordinates reads
\begin{equation}
F^{\mu\nu}=\left(\begin{array}{cccc}
0 & B_{z} & -B_{y} & -\frac{E_{x}}{c}\\
-B_{z} & 0 & B_{x} & -\frac{E_{y}}{c}\\
B_{y} & -B_{x} & 0 & -\frac{E_{z}}{c}\\
\frac{E_{x}}{c} & \frac{E_{y}}{c} & \frac{E_{z}}{c} & 0
\end{array}\right)\label{eq:3-1}
\end{equation}
Also, for the sake of simplicity, a relaxation time model will be
introduced on the right hand side of Boltzmann's equation by assuming
that all details of collisions can be included in one parameter $\tau$
such that 
\begin{equation}
J\left(ff'\right)=-\frac{f-f^{\left(0\right)}}{\tau}\label{eq:d}
\end{equation}
This simple model has been extensively used in order to assess the
structure of the fluxes-forces relations for single component fluids.
For a more precise value of the transport coefficients present in
such relations, the use of the complete kernel is required. Introducing
Eqs. (\ref{eq:d}) and (\ref{eq:3-1}) in Eq. (\ref{eq:1-1}) together
with Chapman-Enskog's hypothesis, namely
\begin{equation}
f=f^{\left(0\right)}+f^{\left(1\right)}\label{eq:6-1}
\end{equation}
one obtains for the non-equilibrium correction to the distribution
function 
\begin{equation}
f^{\left(1\right)}=-\tau\left\{ v^{\alpha}f_{,\alpha}^{\left(0\right)}+\left(\frac{q}{m}v_{\nu}F^{\mu\nu}\right)\frac{\partial f^{\left(0\right)}}{\partial v^{\mu}}\right\} \label{eq:7-2}
\end{equation}
where
\begin{equation}
f^{(0)}=\frac{n}{4\pi c^{3}zK_{2}\left(\frac{1}{z}\right)}e^{\frac{\mathcal{U}_{\beta}v^{\beta}}{zc^{2}}}\label{f1}
\end{equation}
is the local equilibrium solution \cite{de groot6} which leads to
the expression \cite{benedicks} 
\begin{equation}
f^{\left(1\right)}=-\tau\left\{ v^{\alpha}f_{,\alpha}^{\left(0\right)}+\left(\frac{q}{mc^{2}z}F^{\mu\nu}v_{\nu}\mathcal{U}_{\mu}\right)f^{\left(0\right)}\right\} \label{eq:7-1-2}
\end{equation}
In Eq. (\ref{eq:7-1-2}) $K_{n}\left(\frac{1}{z}\right)$ is the $n-\text{th}$
order modified Bessel function of the second kind. On the other hand,
for the first term on the right hand side of Eq. (\ref{eq:7-2}),
the local equilibrium Euler equations need to be introduced in order
to write the time derivatives of the state variables, that arise from
the functional hypothesis $f\left(x^{\nu},v^{\nu}\right)=f\left(v^{\nu},n\left(x^{\nu}\right),T\left(x^{\nu}\right),\mathcal{U}\left(x^{\nu}\right)\right)$,
in terms of the forces. In order to separate time and space derivatives
in such a term we introduce the following decomposition for an arbitrary
reference frame 
\begin{equation}
v^{\alpha}=v^{\beta}h_{\,\,\beta}^{\alpha}-\left(\frac{v^{\beta}\mathcal{U}_{\beta}}{c^{2}}\right)\mathcal{U}^{\alpha}\label{eq:8}
\end{equation}
such that
\begin{equation}
v^{\alpha}f_{,\alpha}^{\left(0\right)}=v^{\beta}h_{\,\,\beta}^{\alpha}f_{,\alpha}^{\left(0\right)}-\left(\frac{v^{\beta}\mathcal{U}_{\beta}}{c^{2}}\right)\mathcal{U}^{\alpha}f_{,\alpha}^{\left(0\right)}\label{eq:12}
\end{equation}
For the first term in Eq. (\ref{eq:12-1}) one has 
\begin{equation}
v^{\beta}h_{\,\,\beta}^{\alpha}f_{,\alpha}^{\left(0\right)}=v^{\beta}h_{\,\,\beta}^{\alpha}f^{\left(0\right)}\left[\frac{n_{,\alpha}}{n}+\frac{T_{,\alpha}}{T}\left(1+\frac{v^{\beta}\mathcal{U}_{\beta}}{zc^{2}}-\frac{\mathcal{G}\left(\frac{1}{z}\right)}{z}\right)+\frac{v_{\beta}}{zc^{2}}\mathcal{U}_{\,\,,\alpha}^{\beta}\right]\label{eq:15}
\end{equation}
where $\mathcal{G}\left(\frac{1}{z}\right)=K_{3}\left(\frac{1}{z}\right)/K_{2}\left(\frac{1}{z}\right)$
and for the second one 
\begin{equation}
\mathcal{U}^{\alpha}f_{,\alpha}^{\left(0\right)}=f^{\left(0\right)}\left[\frac{1}{n}\mathcal{U}^{\alpha}n_{,\alpha}+\frac{1}{T}\left(1+\frac{v^{\beta}\mathcal{U}_{\beta}}{zc^{2}}-\frac{\mathcal{G}\left(\frac{1}{z}\right)}{z}\right)\mathcal{U}^{\alpha}T_{,\alpha}+\frac{v_{\beta}}{zc^{2}}\mathcal{U}^{\alpha}\mathcal{U}_{\,\,,\alpha}^{\beta}\right]\label{eq:16}
\end{equation}
As mentioned above, the proper time derivatives are eliminated intriducing
Euler's equations which are obtained by neglecting dissipative terms
in the transport equations (\ref{eq:17}), (\ref{eq:18-1}) and (\ref{eq:19}),
which are given by
\begin{equation}
\mathcal{U}^{\mu}\frac{\partial n}{\partial x^{\mu}}=-n\frac{\partial\mathcal{U}^{\mu}}{\partial x^{\mu}}\label{d1}
\end{equation}
\begin{equation}
\left(\frac{n\varepsilon}{c^{2}}+\frac{p}{c^{2}}\right)\dot{\mathcal{U}^{\mu}}+p_{,\alpha}h^{\mu\alpha}=qF^{\mu\nu}\mathcal{U}_{\nu}\label{d2}
\end{equation}
\begin{equation}
\mathcal{U}^{\nu}T_{,\nu}=-\frac{p}{nC_{n}}\mathcal{U}_{,\nu}^{\nu}\label{d3}
\end{equation}
where the specific heat at constant number density $C_{n}=\left(\frac{\partial\varepsilon}{\partial T}\right)_{n}$
has been introduced. Introducing Eqs. (\ref{eq:15}-\ref{d3}) in
Eq. (\ref{eq:12-1}) yields the following expression
\begin{align}
v^{\alpha}f_{,\alpha}^{\left(0\right)} & =f^{\left(0\right)}\left\{ v^{\beta}h_{\beta}^{\alpha}\left[\frac{n_{,\alpha}}{n}+\frac{T_{,\alpha}}{T}\left(1+\frac{v^{\beta}\mathcal{U}_{\beta}}{zc^{2}}-\frac{\mathcal{G}\left(\frac{1}{z}\right)}{z}\right)+\frac{v_{\mu}}{zc^{2}}\mathcal{U}_{,\alpha}^{\mu}\right]\right.\nonumber \\
 & -\left.\left(\frac{v^{\beta}\mathcal{U}_{\beta}}{c^{2}}\right)\left[-\mathcal{U}_{,\mu}^{\mu}-\frac{p}{nTC_{n}}\mathcal{U}_{,\nu}^{\nu}\left(1+\frac{v^{\beta}\mathcal{U}_{\beta}}{zc^{2}}-\frac{\mathcal{G}\left(\frac{1}{z}\right)}{z}\right)+\frac{v_{\mu}}{\tilde{\rho}zc^{2}}\left(-p_{,\nu}h^{\mu\nu}+nqF^{\mu\nu}\mathcal{U}_{\nu}\right)\right]\right\} \label{d4}
\end{align}
Notice that the force term in Eq. (\ref{d4}) will compete with the
second term on the right hand side of Eq. (\ref{eq:7-1-2}). Isolating
these two electromagnetic contributions to the deviation from equilibrium
of the distribution function due to them, which we call $f_{EM}^{\left(1\right)}$,
we obtain
\begin{equation}
f_{EM}^{\left(1\right)}=\frac{\tau q}{zmc^{2}}F^{\mu\nu}v_{\mu}\mathcal{U}_{\nu}f^{\left(0\right)}\left\{ \left(\frac{v^{\beta}\mathcal{U}_{\beta}}{c^{2}}\right)\frac{1}{\mathcal{G}\left(\frac{1}{z}\right)}+1\right\} \label{d6}
\end{equation}
where we have used $\tilde{\rho}=nm\mathcal{G}\left(\frac{1}{z}\right)$.
Equation (\ref{d6}) is the main result of this section. It consitutes
the generalization of the result obtained in Ref. \cite{benedicks}
where the fact that a purely electrostatic field can drive a single
charged gas out of equilibrium in a relativistic scenario. In the
present work, the generalization is twofold in the sense that it includes
a magnetic field and is consequently expressed in an arbitrary reference
frame. This expression will be used in the next section in order to
establish the heat flux due to this effect.

\section{Calculation of the heat flux}

As extensively discussed in Refs. \cite{microscopic} and \cite{valdemar},
heat is defined as the transport of internal energy arising from the
chaotic motion of the molecules which is in principle a quantity characteristic
of the comoving frame in which all mechanical effects are absent.
Also, as shown in Ref. \cite{microscopic}, it can be constructied
as a tensor quantity and can thus be calculated in any arbitrary frame.
The expression
\begin{equation}
Q^{\beta}=h_{\eta}^{\beta}\mathcal{L}_{\gamma}^{\eta}\int k^{\gamma}f^{\left(1\right)}\gamma_{\left(k\right)}^{2}d^{*}K\label{eq:heatFlux rel-1-1}
\end{equation}
is here used in order to assess the contribution to thermal dissipation
due to the electromagnetic field \cite{microscopic} in an arbitrary
reference frame in which the hydrodynamic four velocity of the fluid
is given by $\mathcal{U}^{\nu}$. In Eq. (\ref{eq:heatFlux rel-1-1})
$\mathcal{L}_{\gamma}^{\eta}$ is a Lorentz boost with velocity $\vec{u}$
and $K^{\nu}=\gamma_{\left(k\right)}\left[\vec{k},c\right]$ is the
chaotic velocity, that is the particles' velocities measured by an
observer comoving with the fluid's volume element .

In order to calculate the contribution to the integral in Eq. (\ref{eq:heatFlux rel-1-1})
arising from $f_{EM}^{\left(1\right)}$, it needs to be written in
terms of the chaotic velocity. In order to accomplish such task, we
use the invariant expression \cite{valdemar}
\begin{equation}
\frac{v^{\beta}\mathcal{U}_{\beta}}{c^{2}}=-\gamma_{\left(k\right)}\label{eq:s}
\end{equation}
and also write the molecular velocity as
\begin{equation}
v_{\mu}=\mathcal{L}_{\mu\nu}K^{\nu}=\gamma_{\left(k\right)}\mathcal{L}_{\mu\nu}k^{\nu}\label{eq:s1}
\end{equation}
Introducing these expressions in Eq. (\ref{d6}) one obtains
\begin{equation}
f_{EM}^{\left(1\right)}=-\frac{\tau q}{zmc^{2}}\mathcal{L}_{\mu\alpha}F^{\mu\nu}\mathcal{U}_{\nu}\gamma_{\left(k\right)}k^{\alpha}\left\{ \frac{\gamma_{\left(k\right)}}{\mathcal{G}\left(\frac{1}{z}\right)}-1\right\} f^{\left(0\right)}\label{eq:s2}
\end{equation}
Thus, the electromagnetic contribution to the heat flux is given by
\begin{equation}
Q_{EM}^{\beta}=-\frac{\tau q}{z}\mathcal{L}_{\mu\alpha}F^{\mu\nu}\mathcal{U}_{\nu}h_{\eta}^{\beta}\mathcal{L}_{\gamma}^{\eta}\int f^{\left(0\right)}k^{\alpha}k^{\gamma}\left\{ \frac{\gamma_{\left(k\right)}}{\mathcal{G}\left(\frac{1}{z}\right)}-1\right\} \gamma_{\left(k\right)}^{3}d^{*}K\label{eq:heatFlux rel-1-1-1}
\end{equation}
which can also be written as
\begin{equation}
Q_{EM}^{\beta}=-\frac{\tau q}{z}\mathcal{L}_{\mu\alpha}\mathcal{L}_{\gamma}^{\eta}h_{\eta}^{\beta}F^{\mu\nu}\mathcal{U}_{\nu}\mathcal{I}^{\gamma\alpha}\left(z\right)\label{i}
\end{equation}
where the integral $\mathcal{I}^{\alpha\gamma}$ is given by 
\begin{equation}
\mathcal{I}^{\alpha\gamma}\left(z\right)=4\pi c^{3}\int f^{\left(0\right)}k^{\alpha}k^{\gamma}\left\{ \frac{\gamma_{\left(k\right)}}{\mathcal{G}\left(\frac{1}{z}\right)}-1\right\} \gamma_{\left(k\right)}^{3}\sqrt{\gamma_{\left(k\right)}^{2}-1}d\gamma\label{eq:12-1}
\end{equation}
Equation (\ref{i}) contains the main result of this work: the heat
flux for a single component charged fluid in the special relativistic
regime features a contribution due to the electromagnetic force $qF^{\mu\nu}\mathcal{U}_{\nu}$.
This effect is purely relativistic and generalices the result obtained
in Ref. \cite{benedicks} where the particular case of a purely electrostatic
field in a comoving frame was addressed. It is important to emphasize
that the dissipative flux in Eq. (\ref{i}) is a tensor quantity.

In order to assess the magnitude of the corresponding transport coefficient,
the values for the integral $\mathcal{I}^{ab}\left(z\right)$ need
to be obtained. This can be readily done yielding

\[
\mathcal{I}^{ab}\left(z\right)=0,\quad\text{for }a\neq b,\, a,b=1,2,3,4
\]
\[
\mathcal{I}^{aa}\left(z\right)=nzc^{2}\left(5z+\frac{1}{\mathcal{G}\left(\frac{1}{z}\right)}-\mathcal{G}\left(\frac{1}{z}\right)\right)\quad\text{for }a=1,2,3
\]
\[
\mathcal{I}^{44}\left(z\right)=nzc^{2}\left(5z+\frac{2}{3}\frac{1}{\mathcal{G}\left(\frac{1}{z}\right)}-\mathcal{G}\left(\frac{1}{z}\right)\right)
\]
Also notice that the $\gamma=\alpha=4$ term in Eq. (\ref{i}) vanishes
since $\mathcal{L}_{\mu4}\propto\mathcal{U}_{\mu}$ and thus one can
write
\begin{equation}
Q_{EM}^{\beta}=-\kappa_{NR}\left(\frac{q}{kT}\right)\frac{2}{5}\left(5+\frac{1}{z\mathcal{G}\left(\frac{1}{z}\right)}-\frac{\mathcal{G}\left(\frac{1}{z}\right)}{z}\right)\mathcal{L}_{\mu b}\mathcal{L}_{a}^{\eta}h_{\eta}^{\beta}F^{\mu\nu}\mathcal{U}_{\nu}\delta^{ab}\label{i-1-1}
\end{equation}
with $a,b=1,2,3$. The non-relativistic thermal conductivity in the
relaxation time approximation $\kappa_{NR}=\frac{5}{2}\frac{nk^{2}T^{2}}{m}\tau$
has been here introduced. 

The components of the heat flux in Eq. (\ref{i-1-1}) can be explicitly
obtained in an arbitrary frame where $\mathcal{U}^{\nu}=\left[\vec{u},c\right]$,
which yieds:

\begin{equation}
Q_{EM}^{\nu}=-\kappa_{NR}\left(\frac{q}{kT}\right)\frac{2}{5}\left(5+\frac{1}{z\mathcal{G}\left(\frac{1}{z}\right)}-\frac{\mathcal{G}\left(\frac{1}{z}\right)}{z}\right)\gamma_{u}\left[\vec{E}+\vec{u}\times\vec{B},\vec{E}\cdot\frac{\vec{u}}{c}\right]\label{eq:df}
\end{equation}
from which one can extract three relevant limits. Firstly notice that
in the comoving frame where $\mathcal{U}^{\nu}=\left[\vec{0},c\right]$
and $\gamma_{u}=1$ one obtains
\[
Q_{EM}^{\nu}=-\kappa_{NR}\left(\frac{q}{kT}\right)\frac{2}{5}\left(5+\frac{1}{z\mathcal{G}\left(\frac{1}{z}\right)}-\frac{\mathcal{G}\left(\frac{1}{z}\right)}{z}\right)\left[\vec{E},0\right]
\]
which is precisely the result obtained in Ref. \cite{benedicks} for
the case of a purely electrostatic field in the fluid's comoving frame.
The other two limits pertain the non-relativistic case. For a fluid
with a high value of $z$ but such that the hydrodynamic velocity
is considerably lower than the speed of light, $\gamma_{u}\sim1$
and
\begin{equation}
Q_{EM}^{\nu}\sim-\kappa_{NR}\left(\frac{q}{kT}\right)\frac{2}{5}\left(5+\frac{1}{z\mathcal{G}\left(\frac{1}{z}\right)}-\frac{\mathcal{G}\left(\frac{1}{z}\right)}{z}\right)\left[\left(\vec{E}+\vec{u}\times\vec{B}\right),\vec{E}\cdot\frac{\vec{u}}{c}\right]\label{eq:df-1}
\end{equation}
which is physically meaningfull since the thermodynamic force in this
limit is the electromagnetic force for the spatial components and
corresponds to the mechanical electrostatic dissipation on the fourth
component. Finally, the striclty non-relativistic limit in which also
the temperature is low such that $z\ll1$ one has
\begin{equation}
Q_{EM}^{\nu}\sim-\kappa_{NR}\left(\frac{q}{kT}\right)\left(z-z^{2}+....\right)\left[\left(\vec{E}+\vec{u}\times\vec{B}\right),\vec{E}\cdot\frac{\vec{u}}{c}\right]\label{eq:df-1-1}
\end{equation}
finally showing that this effect is strictly relativistic.

\section{Final remarks}

This paper has been devoted to the extension of the Benedicks-type
effect recently identified in the context of special relativity to
the magnetic field case. Although the calculation may seem to be a
straightforward exercise considering the inclusion of the magnetic
field components in Faraday's field tensor when calculating $f^{(1)}$,
there are several features that must be highlighted. The fact that
$\vec{B}$ appears explicitly in $f^{(1)}$ implies a non-zero contribution
of the magnetic field to the entropy production present in a single
component fluid. In contrast, in the non-relativistic case the magnetic
field does not contribute to the entropy production at the Navier-Stokes
level \cite{cc}. More over the hydrodynamic equations now include
new terms in which the field is now identified as a thermodynamic
force. This is rather new, since the only dissipative effects related
to magnetic fields previously identified at first order in the Knudsen
parameter (for simple fluids) were those associated to the values
of the viscosity coefficients \cite{cc,balescu,key-2}.

Magnetic fields are well known to be ubiquous in the universe. Although
the corresponding astrophysical mean free times are quite large compared
to the ordinary scales, the characteristic times for most processes
are such that dissipative effects must be taken into account {[}Spitzer{]}.
It is in this type of scenarios in which irreversible thermodynamics
becomes irrelevant. Future work will be devoted the study of hydromagnetic
instabilities including the effects presented in this work.

\textsf{\textbf{\textit{\Large Acknowledgements}}}{\Large \par}

The authors acknowledge support from CONACyT through grant number
CB2011/167563.

\end{document}